\documentclass[pra,aps,onecolumn,showpacs]{revtex4}
\usepackage{amssymb}
\usepackage{amsmath}
\usepackage[dvips]{graphicx}
\usepackage[english]{babel}

\begin{document}

\title{Vortex nucleation in rotating BEC: the role of the boundary condition
for the order parameter}
\author{W. V. Pogosov and K. Machida}
\affiliation{Department of Physics, Okayama University, Okayama 700-8530, Japan}
\date{\today }

\begin{abstract}
We study the process of vortex nucleation in rotating two-dimensional BEC
confined in a harmonic trap. We show that, within the Gross-Pitaevskii
theory with the boundary condition of vanishing of the order parameter at
infinity, topological defects nucleation occurs via the creation of
vortex-antivortex pairs far from the cloud center, where the modulus of the
order parameter is small. Then, vortices move towards the center of the
cloud and antivortices move in the opposite direction but never disappear.
We also discuss the role of surface modes in this process.
\end{abstract}

\maketitle

\section{Introduction}

Quantized vortices in Bose-Einstein condensates (BEC) of alkali atom gases
attract a considerable current interest. One of the most interesting
problems of vortex physics is the nucleation of topological defects. A lot
of experimental \cite{1,2,3,4} and theoretical \cite%
{5,6,7,8,9,10,11,12,13,14,15,16,17,18,19,20,21} works deal with this
subject. Phases with different number of vortices are usually separated by
large energetic barriers. Therefore, the transition to the vortex state
occurs at higher rotation frequency than the thermodynamical one determined
by the balance of energies of vortex-free and one-vortex states. The same is
valid for the well-studied case of superconductors \cite{23,24}. However,
BEC clouds have no well-defined surface and, therefore, the process of
vortex formation must be different from that in superconductors, where
vortex lines penetrate from the sample surface. Within the Gross-Pitaevskii
theory, the boundary condition for the order parameter in the harmonically
trapped BEC reads%
\begin{equation}
\psi (r\rightarrow \infty )=0.  \label{BoundaryCondition}
\end{equation}

In the li1 of strongly interacting gas, a Thomas-Fermi approximation is
often used. In this approach, the Thomas-Fermi boundary, where $\psi $
vanishes identically, has to be assumed, and vortices are usually treated as
point-like objects. Under these assumptions, one can calculate the critical
rotation frequency, which suppresses the surface barrier preventing vortex
penetration through the Thomas-Fermi boundary \cite{11,12,7}. Another
approach to the problem \cite{5,9,10,13} is based on the analysis of
stability of the \textit{axially-symmetric} vortex-free state with respect
to the perturbations of the order parameter $\psi $, and it was shown that
the nonvortex state becomes unstable with respect to surface modes, which
leads to the well-known Landau instability. It is believed that this
instability results in the nucleation of vortices. The third approach to the
problem of vortex formation is by numerical solutions of the time-dependent
Gross-Pitaevskii (GP) equation \cite{15,14,16,6,8}, supplemented by the
boundary condition of varnishing of $\psi $ at some boundary.

The aim of the present paper is to study a vortex nucleation process taking
into account consistently boundary condition (1) for the order parameter. We
consider the case, when the system is spined up quasistatically and
adiabatically. No boundary, which can serve as a gate for vortex penetration
is assumed. We use a semi-analytical model \cite{25}, which is applicable
for the li1 of diluted gas. We show that nucleation of vortices in the model
with boundary condition (1) occurs via the creation of vortex-antivortex
pairs far from the cloud center, in regions, where $\left\vert \psi
\right\vert $\ is small. Vortices then move toward the cloud center to
decrease the energy of the system, whereas antivortices go to the opposite
direction. This mechanism for the vortex nucleation is natural for the case
of a system, which has no boundary and where the total topological charge
has to be conserved. Also studied is the role of surface modes in the
process of vortex nucleation. We show that the critical velocity for vortex
formation is slightly higher than the Landau critical velocity, and this is
in a agreement with Anglin's results \cite{11,12} obtained by the very
different method.

\section{Model}

We consider the case of quasi two-dimensional condensate. The dimensionless
GP energy functional is given by:

\begin{equation}
F=\int rdr\int d\varphi \Big(\frac{1}{2}|\nabla \psi |^{2}+\frac{r^{2}}{2}%
|\psi |^{2}+\frac{2\pi p}{N}|\psi |^{4}+i\omega \psi ^{\ast }\;\frac{%
\partial \psi }{\partial \phi }\Big),  \label{energi}
\end{equation}%
where the integration is performed over the area of the system, $(r,\varphi
) $ are polar coordinates, $N$ is the number of atoms, $\omega $ is the
rotation frequency, $p=an_{z}$ is the gas parameter, $a$ and $n_{z}$ are
scattering length and concentration of particles along the cloud axis,
respectively. Distances and rotation frequencies are measured in units of
the oscillator length and the trapping frequency, respectively. The
normalization condition for the order parameter reads $\int rdr\int d\varphi
\left\vert \psi \right\vert ^{2}=N$.

\textit{In general case}, $\psi $ can be represented as a Fourier expansion%
\begin{equation}
\psi (r,\varphi )=\sum_{l}f_{l}(r) \exp (-il\varphi ).  \label{Fur'e}
\end{equation}%
In the li1 of noninteracting gas ($p=0$), it can be easily shown by
substitution of Eq. (3) to the GP equation that each function $f_{l}(r)$
coincides with the eigen function for the harmonic oscillator corresponding
to the angular momentum $l$. These functions have the Gaussian profile $\sim
r^{l}\exp \left( -\frac{r^{2}}{2}\right) $. Therefore, one can assume that
this Gaussian approximation remains accurate in the case of weakly
interacting dilute gas. The accuracy can be improved if we introduce a
variational parameter $R_{l}$ characterizing the spatial extend of $f_{l}(r)$%
. Finally, our \textit{ansatz} for $f_{l}(r)$ has a form: 
\begin{equation}
f_{l}(r)=C_{l}\left( \frac{r}{R_{l}}\right) ^{l}\exp \left( -\frac{r^{2}}{%
2R_{l}^{2}}-i\phi _{l}\right) ,
\end{equation}%
where $C_{l}$, $R_{l}$, and $\phi _{l}$ can be found from the condition of
minimum of the energy (2), $C_{l}$ is a real number. This approach was used
for the first time in Ref. \cite{25}, see also Ref. \cite{26,27}. In the
present paper, we restrict ourselves to the interval $0<p<10$, since,
according to our estimates, the Gaussian approximation becomes inaccurate at
higher $p$. To verify this, we compared known numerical results for the
thermodynamical critical rotation frequency with those obtained by our
method.

Now we substitute Eqs. (3) and (4) to Eq. (2) and after the integration we
obtain:

\ 
\begin{eqnarray}
F &=&\sum_{l}\alpha
_{l}C_{l}^{2}+\sum_{l}I_{llll}C_{l}^{4}+4\sum_{l>k}I_{llkk}C_{l}^{2}C_{k}^{2}
\notag \\
&&+4\sum_{l>k>m}I_{lkkm}C_{l}C_{k}^{2}C_{m}\delta _{l+m,2k}\cos \left( \phi
_{l}+\phi _{m}-2\phi _{k}\right) \smallskip  \notag \\
&&+8\sum_{l>k>m>n}I_{lkmn}C_{l}C_{k}C_{m}C_{n}\delta _{l+k,m+n}\cos \left(
\phi _{l}+\phi _{k}-\phi _{m}-\phi _{n}\right) ,  \label{ener}
\end{eqnarray}%
where 
\begin{equation}
\alpha _{l}=\frac{\pi }{2}\Gamma (l+2)\left( 1+R_{l}^{4}\right) +\pi
R_{l}^{2}\Gamma (l+1)\omega l,  \label{alfa}
\end{equation}%
\begin{eqnarray}
I_{lkmn} &=&\frac{2\pi ^{2}p}{N}\Gamma (\frac{l+m+n+k}{2}+1)R_{lkmn}^{2} 
\notag \\
&&\times \left( \frac{R_{lkmn}}{R_{l}}\right) ^{l}\left( \frac{R_{lkmn}}{%
R_{k}}\right) ^{k}\times \left( \frac{R_{lkmn}}{R_{m}}\right) ^{m}\left( 
\frac{R_{lkmn}}{R_{n}}\right) ^{n},  \label{integ}
\end{eqnarray}%
\begin{equation}
R_{lkmn}=\sqrt{2}\left( R_{l}^{-2}+R_{k}^{-2}+R_{m}^{-2}+R_{n}^{-2}\right)
^{-\frac{1}{2}},  \label{radius}
\end{equation}%
$\Gamma (l)$ is a gamma function. Normalization condition is now given by 
\begin{equation}
\pi \sum_{l}C_{l}^{2}R_{l}^{2}\Gamma (l+1)=N.  \label{normalization}
\end{equation}%
Values of parameters $R_{l}$, $C_{l}$\ and $\phi _{l}$\ can be found from
the minimum of the energy (5) taking into account Eq. (9). For instance, for
the axially-symmetric vortex-free state, $C_{0}=\sqrt{N/\pi R_{0}^{2}}$, $%
R_{0}=\left( 1+2p\right) ^{1/4}$, and $C_{l}=0$ at $l\eqslantgtr 1$. \ 

In the vicinity of the local minimum, the deviation of the energy from the
minimum can be represented as

\begin{equation}
\Delta F=\sum_{lk}A_{lk}\Delta C_{l}\Delta C_{k},  \label{stability}
\end{equation}%
where $\Delta C_{l}$\ is a deviation of $C_{l}$\ from its equilibrium value.
In Eq. (10) we kept the terms up to the second order of $\Delta C_{l}$. The
stability criterion of the solution is a positiveness of all main minors of
the matrix $\left\Vert A_{lk}\right\Vert $.

\section{Stability of nonvortex state}

First, we study the stability of the vortex-free state and after this we
will go beyond this linear approach. We consider the situation, when the
system is spined up quasistatically. For the axially-symmetric nonvortex
state, the matrix $\left\Vert A_{lk}\right\Vert $ is diagonal. In this case,
it is easy to find $\omega $, at which vortex-free phase becomes unstable
with respect to the surface mode with the angular momentum $l$:

\begin{eqnarray}
\omega _{inst}(l,p) &=&\frac{2}{l\Gamma (l+1)}\left[ \frac{(l-1)\Gamma (l+1)%
}{\sqrt{1+2p}}+\right.  \notag \\
&&\frac{1}{4}\left[ \frac{1+R_{l}^{4}}{R_{l}^{2}}\right] \Gamma (l+2)+\frac{%
4(1+2p)^{l/2}\Gamma (l+1)}{(\sqrt{1+2p}+R_{l}^{2})^{l+1}}.  \label{frequency}
\end{eqnarray}%
Value of $R_{l}$ can be found from the condition of minimum of $\omega
_{inst}(l,p)$ with respect to $R_{l}$. In Fig. 1 (a) we plot the dependences
of $\omega _{inst}(l,p)$ on $p$ for different $l$. The resulting function $%
\omega _{inst}(p)$ is determined by the minimal value of $\omega
_{inst}(l,p) $ with respect to the quantized $l$.\ This function is shown in
Fig. 1 (b) and represents a well-known Landau criterion for the
angular-momentum transfer to the superfluid system. From Fig. 1 (b) one can
see that value of $l_{inst}$, which leads to the instability, increases with
increase of gas parameter $p$. Vortex-free phase is always stable with
respect to the generation of harmonic $l=1$. At low $p$ values, the
instability with respect to the harmonic $l=3$ occurs at lower $\omega $
than that for $l=2$. In Fig. 1 (b), we also presented the thermodynamical
critical rotation frequency as a function of $p$, which we calculated by the
comparison of energies of vortex-free and single-vortex states. It is much
lower than $\omega _{inst}$, as can be expected.

\section{Nucleation of vortices}

Now we go beyond the linear analysis. It follows from the results of the
previous Section that, in the instability point, an admixture of a harmonic
with $l=l_{inst}$ appears in the order parameter expansion (3). In fact, the
order parameter must contain also all other harmonics with $l$'s divisible
by $l_{inst}$, which are induced by the component with $l=l_{inst}$, as can
be easily seen from the GP equation. However, the main contribution to the
energy is given by just two first terms with $l=0$ and $l=l_{inst}$.
Therefore, we will take into account only these two harmonics. The component
with $l=l_{inst}$ gives $l_{inst}$-fold modulation of the modulus of the
order parameter in the azimuthal direction and it is responsible for the
creation of vortices. We calculate the energy and $\psi $ in the
two-harmonic approximation at values of $\omega $\ slightly exceeding\ $%
\omega _{inst}$ and study their evolution with changing of $\omega $. At
each $\omega $\ we also check the stability of the obtained solution with
respect to the nucleation of other harmonics with various $l$'s, not
divisible by $l_{inst}$. The two-harmonic approximation was used widely for
the analysis of vortex states in mesoscopic superconducting discs within the
Ginzburg-Landau theory, see e.g. Refs. \cite{28,29,30,31}. Note that we also
tried to take into account several additional components of $\psi $ with $l$%
's divisible by $l_{inst}$ ($2l_{inst}$, $3l_{inst}$, $4l_{inst}$), but this
does not change our results significantly thus verifying good accuracy of
the two-harmonic approximation in our case.

First, we discuss the simplest case of $l_{inst}=3$. In the instability
point, an admixture of harmonic with $l=3$ appears. Surface mode leads to
the three-fold modulation of $\left\vert \psi \right\vert $ in the azimuthal
direction. In Fig. 2 we show the evolution of $\left\vert \psi \right\vert $
far from the system center along one of three directions, where $\left\vert
\psi \right\vert $\ is minimal, at $p=1$. In the beginning, $\left\vert \psi
\right\vert $ in this direction is a monotonic function of $r$. With further
increasing of $\omega $, a local minimum appears at the dependence of $\psi
(r)$, see Fig. 2 (a). This minimum becomes deeper and deeper and then $%
\left\vert \psi \right\vert $ vanishes at some point, as shown on Fig. 2
(b). If we proceed with spinning up the system, the minimum splits to the
vortex-antivortex pair, see Fig. 2 (c). If we go around the center of the
vortex, the phase of the order parameter changes by $2\pi $, and around
antivortex it changes by $-2\pi $. Thus, we can conclude that these zeros of
the order parameter indeed correspond to the vortex and antivortex.
Increasing of $\omega $ leads to the motion of vortex toward the system
center and antivortex in the opposite direction, where $\left\vert \psi
\right\vert $ is very small. The same is happening along the other two
directions. As a result, three vortices penetrate the inner part of a cloud
in a symmetrical pattern, whereas three antivortices move far from the cloud
center but never disappear. The spot, where vortex-antivortex pair
nucleates, is characterized by a small value of $\left\vert \psi \right\vert 
$, approximately, fifty times smaller than $\left\vert \psi \right\vert $ in
the cloud center. Therefore, the formation of pairs is invisible on the
large length-scale. Also, the interval of $\omega $ between the point of the
instability and the vortex state is very narrow, of the order of $10^{-3}$
of trapping frequency. This interval is occupied by the vortex-free state
with the order parameter modulated in the azimuthal direction. We believe
that vortex-antivortex pairs are not an artifact of our method, since they
appear even in the li1 $p<<1$, where Gaussian approximation is very
accurate. Also, this mechanism of vortex nucleation is rather natural for
the continuos system without any borders, as reflected by the boundary
condition (1) for the order parameter.

The similar scenario for the vortex nucleation is realized at values of $p$
corresponding to $l_{inst}=5$. In this case, five vortices penetrate the
inner part of the cloud and five antivortices move in the opposite
direction. Thus, surface modes induce ripples of the order parameter and
then ripples are naturally transformed to the same number of
vortex-antivortex pairs. However, when $l_{inst}$ is even, $l_{inst}=4$ or $%
6 $, the scenario is different. At first, $\left\vert \psi \right\vert $
becomes modulated in the azimuthal direction and with increasing of $\omega $%
, the minima of $\psi $ become more and more pronounced exactly as at odd $%
l_{inst}$. But just before the formation of $l_{inst}$\ vortices this phase
suddenly becomes unstable with respect to the nucleation of harmonic with $%
l=l_{inst}/2$. The stationary state after this instability corresponds to
the local minimum of energy with $\psi $, which is given by a superposition
of harmonics with $l=0$, $l_{inst}/2$, and $l_{inst}$. At $1\lesssim
p\lesssim 1.5$ ($l_{inst}=4$), this phase is also unstable with respect to
the nucleation of the harmonic with $l=1$. This implies that the final state
contains one vortex in the center of the cloud and one antivortex at
infinity. Thus, according to our model, for even values of $l_{inst}$,\ not
all ripples of $\psi $ are transformed to vortex-antivortex pairs. Number of
these pairs is equal to $3$ for $l_{inst}=6$ and $1$ or $2$ for $l_{inst}=4$
depending on $p$. The role of surface modes in this case is to facilitate
the penetration of one or more number of vortices to the inner part of the
cloud. The similar mechanism for vortex penetration was analyzed in Ref. 
\cite{7} \ for the particular case of a quadrupole mode, $l=2$, which
facilitates penetration of a single vortex. No antivortices were found in
this case, since the Thomas-Fermi boundary was assumed. Our result is more
general. It shows that penetration of vortices can be related to the
induction of different surface modes depending on $p$. After the cascade of
various harmonics generation, multiple-vortex penetration can occur, not
necessarily one vortex enters the cloud. In our model, the barrier
preventing the separation of vortex and antivortex (or surface barrier in
models assuming Thomas-Fermi boundary) is destroyed at critical frequencies
slightly exceeding the Landau critical frequency for the surface modes.
Thus, our model supports the Anglin's results \cite{11,12}, which were
obtained by using completely different method, namely, the hydrodynamic
approach taking into account spatial variation of the order parameter in the
Thomas-Fermi surface layer.

The mechanism for the vortex nucleation via vortex-antivortex pairs is a
specific feature of infinite systems. One can see here some analogy with the
BKT transition, where topological defects also nucleate by pairs, and the
total topological charge is always zero. In finite systems, topological
defects usually penetrate through the surface, but here no surface is
assumed, since condition (1) should be fulfilled. Therefore, a natural way
to create vortices is via vortex-antivortex pairs, after which antivortices
are pushed to the regions with small $\left\vert \psi \right\vert $, since
their presence in regions with high $\left\vert \psi \right\vert $ is
energetically unfavorable. Therefore, antivortices are practically
undetectable experimentally. An important consequence of this result is that
the topological charge of BEC with boundary condition (1) is always equal to
zero. This means that for any stable configuration of vortices in the inner
part of the cloud there is a configuration of antivortices at the periphery
of the system, where the d2ity of particles is very low. Indeed, using our
method we calculated $\psi (r,\varphi )$ for different numbers of vortices
and we always found the same number of antivortices in the system. Note that
some additional topological defects in regions of small $\left\vert \psi
\right\vert $ were reported in Refs. \cite{14,15,16}, where int2ive
numerical methods were used for the solution of the GP equation. They were
called \textquotedblleft ghost vortices\textquotedblright . The
correspondence between the antivortices in our model and \textquotedblleft
ghost vortices\textquotedblright\ is an open question.

It can be also predicted that in 3D case vortex loops nucleate in rotating
BEC instead of vortex-antivortex pairs. With increasing $\omega $, lateral
dim2ions of loops grow up and finally 'vortex' parts of loops penetrate the
inner part of cloud and 'antivortex' parts go to the opposite direction.
This effect is related to the bending of vortex lines found in 3D numerical
simulations \cite{22,16}. It is also interesting to analyze the stability of
knotted GP\ solutions in this case. In 3D superconductors, vortices also
penetrate through the surface not as straight lines, but as half-loops \cite%
{24,32}. Physically, this is due to the fact that the vortex energy is
proportional to its length and therefore a creation of vortex from the point
nucleus is easier than from the line one. BKT transition in 3D case also
occurs via the generation of vortex loops \cite{33}, see also \cite{34}.

Now we briefly discuss the case of finite cloud. We perform the same
calculations, but using functions

\begin{equation}
f_{l}(r)=C_{l}\left[ \left( \frac{r}{R_{l}}\right) ^{l}+\beta _{l}\left( 
\frac{r}{R_{l}}\right) ^{l+2}\right] \exp \left( -\frac{r^{2}}{2R_{l}^{2}}%
-i\phi _{l}\right) ,  \label{ModifFunc}
\end{equation}%
where $\beta _{l}$ is an additional parameter, which is chosen in such a way
as to meet the boundary condition $f_{l}(R_{b})=0$, $R_{b}$ is the radius of
the system. We found that topological defects nucleate via vortex-antivortex
pairs, if $R_{b}$ is approximately two times larger than the Thomas-Fermi
radius. With further spinning up of the system, some of antivortices leave
the system and the topological charge is not necessarily zero. When $R_{b}$
is less, vortices penetrate via the surface as usual, and no antivortices
appear.

\section{Conclusions}

We studied a nucleation of topological defects in rotating BEC in 2D case
within the Gross-Pitaevskii theory with boundary condition of varnishing of
the order parameter in infinity. We found that in this model, with
increasing slowly of the rotation frequency, vortex-antivortex pairs appear
in the system. Vortices move to the inner part of the cloud and antivortices
are pushed to the opposite direction. A number of pairs depends on the gas
parameter. Topological charge of the system is always equal to zero, and for
any configuration of vortices there is a configuration of antivortices far
from the cloud center. For 3D case, we predict that vortex loops nucleate
instead of vortex-antivortex pairs. Also studied is the role of surface
modes in the process of vortex formation. We revealed two scenarios: either
all the ripples of the order parameter induced by surface modes are
transformed to the vortices or surface modes facilitate penetration of
vortices through the cascade of various harmonics generation. A multiple
penetration of vortices is possible. In the finite-size clouds, vortices can
either penetrate through the surface or nucleate via vortex-antivortex pairs
depending on the system size. Vortex nucleation occurs at critical
velocities slightly exceeding the Landau critical velocity for the surface
modes.

\section*{Acknowledgements}

Authors acknowledge useful discussions with Y. Castin, M. Tsubota, T. K.
Ghosh, T. Mizushima, and S. Ghosh. W. V. Pogosov is supported by the Japan
Society for the Promotion of Science. 

\section{Figures}

Fig. 1. The dependences of rotation frequences $\omega _{inst}(l)$, at which
the axially-symmetric nonvortex state becomes unstable, on gas parameter $p$
at different values of quantized angular momentum $l$ (Fig. 1 (a)). Fig. 1
(b) shows the resulting function, which is given by minimum of $\omega
_{inst}(l)$ with respect to $l$. Dot lines on Fig. 1 (b) are giudes for eyes
indicating the boundaries between regions corresponding to different values
of $l_{inst}$. Dashed line on Fig. 1 (b) is the thermodynamical critical
frequency.

Fig. 2. Evolution of the modulus of the order parameter in the vicinity of
the spot, where vortex-antivortex pair nucleates. Fig. 2 (a) corresponds to
the rotation speed 0.77405, Fig. 2 (b) to 0.77417, and Fig. 2 (c) to
0.77419. Number of particles is 10000, $p=1$. 

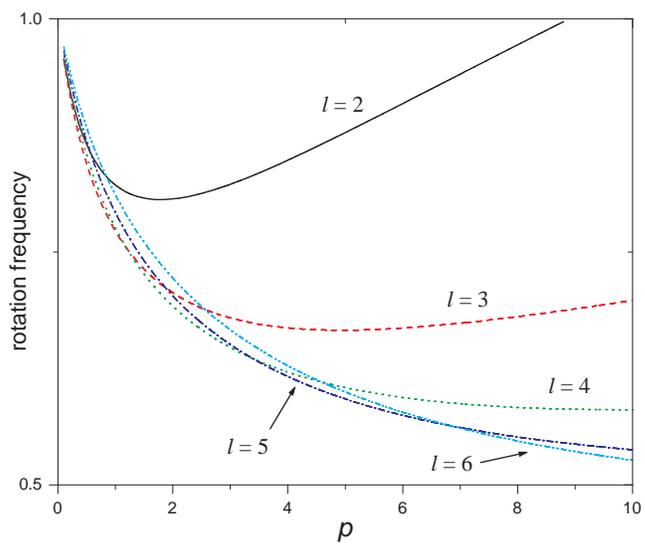
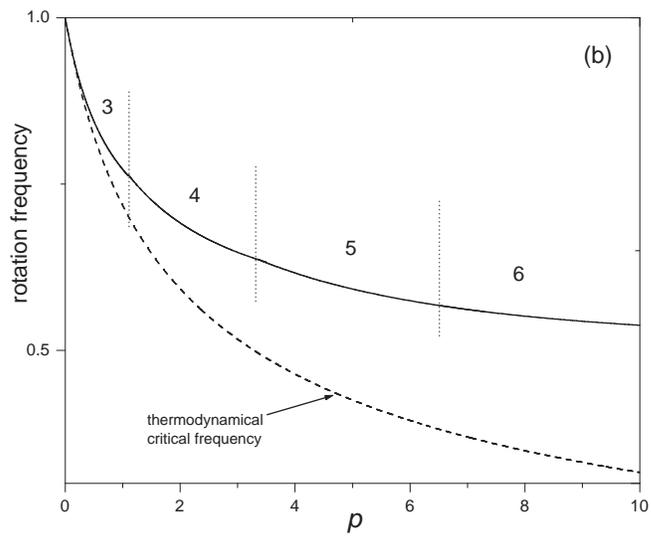
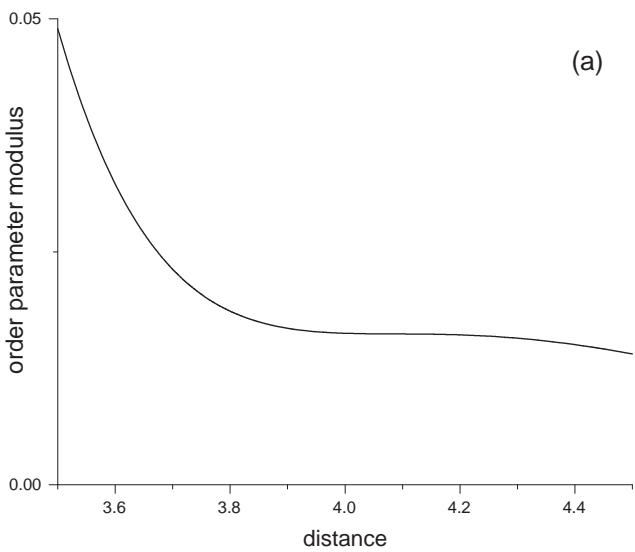
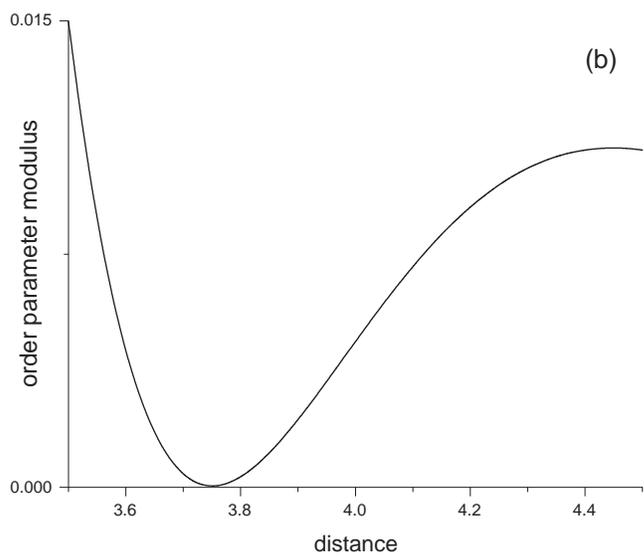
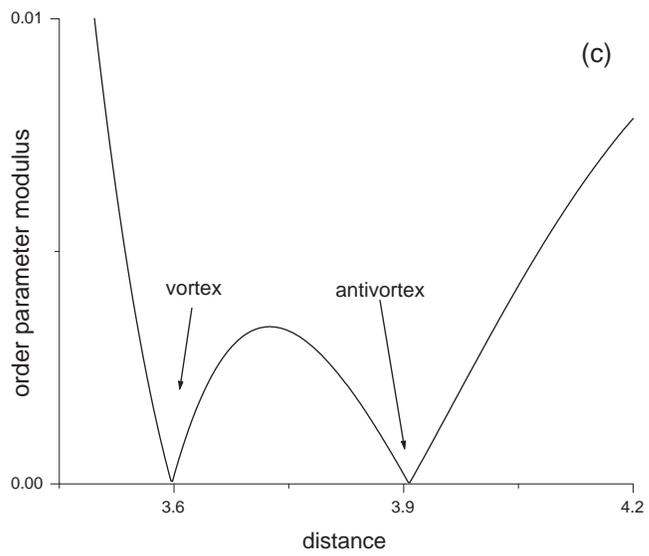
\end{document}